\documentstyle[aps,twocolumn,pre,epsf]{revtex}

\newcommand{\lb}{\overline{\ell}}
\newcommand{\lbsq}{\overline{\ell^2}}
\newcommand{\ra}{\rangle}
\newcommand{\la}{\langle}
\begin{document}
\twocolumn[\hsize\textwidth\columnwidth\hsize\csname
@twocolumnfalse\endcsname
\title{Characterizing the structure of small-world networks}
\author{E. Almaas$^{1,}$\cite{e-mail}, R. V. Kulkarni$^{2}$ and
D. Stroud$^1$}
\address{$^1$Department of Physics, The Ohio State University,
Columbus, OH 43210\\ $^2$Department of Physics, University of
California, Davis}
\date{\today}

\maketitle

\begin{abstract}
We give exact relations which are valid for small-world networks
(SWN's) with a general `degree distribution', i.e the distribution
of nearest-neighbor connections. For the original SWN model, we
illustrate how these exact relations can be used to obtain
approximations for the corresponding basic probability distribution.
In the limit of large system sizes and small disorder, we use
numerical studies to obtain a functional fit for this distribution.
Finally, we obtain the scaling properties for the mean-square
displacement of a random walker, which are determined by the scaling
behavior of the underlying SWN.
\end{abstract}

\draft
\vskip1.5pc]

Networks occur in many contexts in the sciences and humanities. Among
these are neural networks \cite{watts98}, social networks
\cite{watts98,bernard98,newman01}, food webs
\cite{cohen90,williams00}, and computer networks
\cite{faloutsos99,huberman99_1,huberman99_2}. Many properties of such
complex systems can be understood by considering the network of
interactions which connects their components. For this reason, much
recent research has focused on the structure of these networks
\cite{newman99,barabasi99,barabasi99_2,newman99_2,newman00,kulkarni00}.

The study of network models has recently been advanced by the
introduction of small-world networks (SWN's) by Watts and Strogatz
\cite{watts98}. These networks have an ordered structure locally, but
are random on a global scale. This combination of features suggests
that SWN's can be used to describe the behavior of many real-world
interacting networks\cite{strogatz01}. However, as originally
defined, SWN's do not exhibit a property found in some large-scale
networks such as the world-wide web, in which the probability
distribution for the number of nearest-neighbor connections to a given
site (the `degree distribution') is a power law, rather than
Poisson-like, as in the SWN's \cite{barabasi99}. Alternative models
which do exhibit this power-law distribution have been proposed
\cite{albert01,dorogovtsev01,newman01_2,puniyani01}.

In this article, we study SWN's where shortcuts according to a given
degree distribution are superimposed on a regular network. For this
class of networks, we show some exact results which are {\em
independent} of the specific degree distribution employed. These can
be used to find approximations for the basic probability distribution
for the network. Furthermore, on the basis of numerical simulations of
the original SWN model (with Poisson-like degree distribution), we
propose that the basic probability distribution is characterized by a
function of a single variable in the limit of small disorder (as
defined below) and large system size: the so-called `small-world'
regime. Using this functional form, various structural properties of
networks in the small-world regime can be accurately determined
without the need for further, expensive computer simulations. We also
study how the structural properties of this particular SWN model
affects an important example of dynamics on SWN's: the time-dependent
mean-square displacement of a random walker.

The SWN's are defined by starting from a one-dimensional regular
network with periodic boundary conditions and $L=2 N$ nodes, each node
being connected to its $2k$ nearest neighbors. We add shortcut ends to
each node according to a given degree distribution by following the
prescription of Ref.\ \cite{newman01_2}. For the original SWN model,
the degree distribution is ${\cal D}_q = (1-p) ~{\cal P}_q (k p) + p~
{\cal P}_{(q-1)} (k p)$, where ${\cal P}_q (\lambda)= \exp(-\lambda)
\lambda^q /q!$ is the Poisson distribution. We then select pairs of
shortcut ends at random and connect them to each other, thus creating
a shortcut. This procedure (with the above ${\cal D}_q$) of generating
the networks is equivalent to the procedure outlined by Newman and
Watts \cite{newman99}, with $k p$ as the probability of having a
shortcut at a given site. Thus, on average, there will be $x = k p L$
shortcuts in the network. The present numerical results were
calculated for $k=1$; however, generalization to other $k$ is
straightforward and indicated in our equations (which we have verified
numerically). The `small disorder' regime corresponds to $p \ll 1$,
and we will work in the limit of large system sizes such that terms of
${\cal O}(1/L)$ can be omitted.

The SWN is characterized by two types of distances: the `Euclidean
distance,' defined as the shortest distance between two sites before
any shortcuts are introduced into the network; and the `minimal
distance,' which denotes the shortest distance after shortcuts have
been added. An important probability distribution in SWN's is then
$P(m|n)$, defined as the probability that two sites have minimal
separation $m$, given that their Euclidean distance is $n$. As shown
previously\cite{kulkarni00}, $P(m|n)$ can be written in the form
\begin{equation}
P(m|n) = \left\{ \begin{array}{cl}
		f(m), & m < n,\\
	 	1-\sum_{i=1}^{n-1}f(i), & m=n, \\
                0, & m > n. 
		 \end{array} \label{eq:pmn}
         \right.
\end{equation}
Thus, $P(m|n)$ is fully characterized by the function $f(m)$
\cite{fmnote} which can be regarded as the probability that two
diametrically opposite sites have minimal separation $m$. The above
result is {\em exact} for SWN's with arbitrary degree distributions.

Many structural properties, such as the average minimal separation
between two randomly chosen points, $\lb$, can be expressed in terms
of $f(m)$\cite{kulkarni00}, and these expressions can be used to
derive ${\it exact}$ relations connecting various quantities of
interest. An important relation which can be derived in this manner,
relates the average number of sites within $m$ hops of a randomly
selected origin, $V(m)$, to $F(m) = \sum_{k=1}^m f(k)$, the cumulative
probability distribution, as
\begin{equation}
V(m) = 2 k m + \big[L - 2 k m \big] F(m) \label{eq:vm},
\end{equation}
Through our simulations, we have confirmed the validity of
Eqs. (\ref{eq:pmn}) and (\ref{eq:vm}) for several different types of
degree distributions, including networks with power-law degree
distributions. These exact relations can be used to obtain quantities
of interest for the network. In a simple model for disease-spreading
\cite{newman99_2}, for example, $V(m)$ corresponds to the number of
infected sites after $m$ time steps. Furthermore, Eq.\ (\ref{eq:vm})
tells us how to calculate $F(m)$ [and hence $f(m)$], once we know
$V(m)$. Thus, any approximation for $V(m)$ can be converted into an
approximation for $f(m)$. In the remainder of this article, we will
illustrate this procedure by focusing on the original SWN model.

We will now use the mean-field result of Newman {\it et. al} for
$V(m)$ to obtain an approximation for $f(m)$, and hence, to motivate a
new scaling form for this function. In the asymptotic limit ($x \gg
1$), using Eq.\ (\ref{eq:vm}) in conjunction with the mean-field
results of Ref.\ \cite{newman00}, we obtain a relation between $f(m)$
and $F(m)$:
\begin{equation}
f_{\mathrm{mft}}(m) = 4~k^2p ~F_{\mathrm{mft}}(m)~ (1 -
                      F_{\mathrm{mft}}(m)),\label{eq:log}
\end{equation}
which is the logistic growth equation, as pointed out by Barbour {\it
et al} \cite{barbour00}. Eq.\ (\ref{eq:log}) can easily be solved for
$f_{\mathrm{mft}}(m)$\cite{johnson95}. We note that the final
expression for $f_{\mathrm{mft}}(m)$ is consistent with the scaling
form $ f(m;N,p) = h(m/N,pN)/N$ derived in Ref. \cite{kulkarni00}.

We now consider the central moments of this distribution, defined as
\begin{equation}
\mu_\ell = \sum_{m=1}^N \big[m - \la m \ra \big]^\ell
           f_{\mathrm{mft}}(m). \label{eq:mu}
\end{equation}
The scaling form for $f(m)$ implies that the scaled moments
$\mu_\ell/N^\ell$ are functions of $x$ only; however,
$f_{\mathrm{mft}}(m)$ has an additional property: its central moments
in this case take the form
\begin{equation}
\frac{\mu_{\ell}}{N^\ell} = \frac{c_\ell}{x^\ell},
          \label{eq:muk}
\end{equation} 
where $c_\ell$ is the $\ell^{\mathrm{th}}$ central moment of the
logistic distribution. We have confirmed through numerical simulations
that Eq.\ (\ref{eq:muk}) holds for at least the first four central
moments of $f(m)$, however, with coefficients $c_\ell$ which differ
significantly from those obtained from the logistic growth
\begin{figure}[tb]
\epsfysize=7cm
\centerline{\epsffile{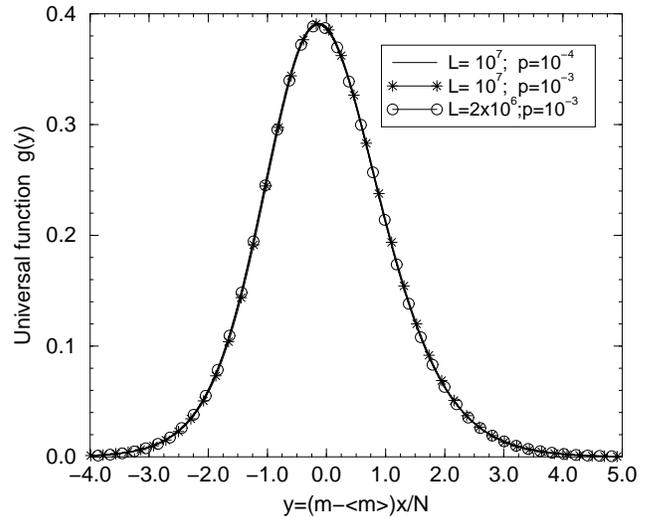}}
\caption{Plot of the universal function $g(y)$ [cf.\ Eq.\
         (\ref{eq:scale})] for three choices of system size $L$ and
         probability of shortcuts, $p$ (see text): (i) $L = 10^7$, $p
         = 10^{-4}$ (full line); (ii) $L=10^7$, $p=10^{-3}$ (stars and
         full line); and (iii) $L=2\times 10^6$, $p=10^{-3}$ (circles
         and full line).}
\label{fig:collapse}
\end{figure} \noindent
equation. For example, from our simulations, we find $c_2 \approx
1.21$, whereas the corresponding mean-field value is
$c_2^{\mathrm{mft}} \approx 0.82$. Eq.\ (\ref{eq:muk}) also implies
that $ \mu_\ell \sim p^{-\ell}$, and hence, for fixed $p$, is
independent of $N$. Consequently, the probability distributions
$f(m)$, for fixed $p$ and different values of $N$, are simply related
by a translation of the mean values. We have confirmed this
translation property by simulations carried out for a wide range of
parameters in the small-world regime.

On the basis of these features, we propose the following scaling form
for $f(m)$ in the asymptotic limit ($x \gg 1$):
\begin{equation}
f(m;N,p) ~=~ \frac{kx}{N}~g\left( \frac{kx}{N}
   (m - \la m \ra) \right), \label{eq:scale}
\end{equation}
where $g(y)$ is a universal function. It is readily seen that the
above scaling form is consistent with Eqn. (\ref{eq:muk}). In Fig.\
\ref{fig:collapse}, we show calculations of $f(m)$ for three different
combinations of the parameters $(p,L)$: $(10^{-4},10^7)$,
$(10^{-3},10^7)$, and $(10^{-3},10^6)$. For all three choices of
parameters, $f(m;N,p)$ can be collapsed onto a single curve. Thus,
numerical simulations for just one choice of the parameters $(p,L)$
suffice to determine $g(y)$. To fully determine $f(m)$ we need to find
functional forms for both $g(y)$ and $\la m \ra$, which we now proceed
to do.

\begin{figure}[tb]
\epsfysize=7cm
\centerline{\epsffile{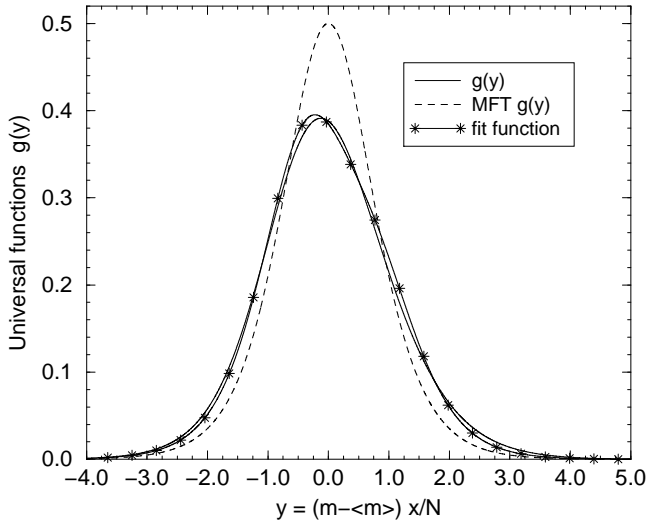}}
\caption{Comparison of the universal function $g(y)$, as obtained from
         numerical calculations (full line), mean-field theory (dashed
         line), and our two-parameter fitting function Eq.\
         (\ref{eq:gfit}) (stars and full line).}
\label{fig:comparison}
\end{figure}

We can represent $g(y)$ using a two-parameter mixture of logistic
distributions:
\begin{equation}
g(y) = \frac{1}{2(1+\gamma)}\bigg[ \mbox{sech}^2(y+\gamma \beta) + \gamma
	~\mbox{sech}^2 (y-\beta) \bigg]. \label{eq:gfit}
\end{equation}
This form ensures that $g(y)$ is normalized, has zero mean, and has
the correct behavior for large $|y|$. In Fig.\ \ref{fig:comparison},
we show a comparison of the mean-field result, the fitting function,
[Eq.\ (\ref{eq:gfit}) with $\gamma = 0.492$ and $\beta = 0.782$], and
the numerically obtained $g(y)$. Evidently, this function 
\noindent
describes the numerical $g(y)$ very well, while the mean-field $g(y)$
agrees less well.

\begin{figure}[tb]
\epsfysize=7cm
\centerline{\epsffile{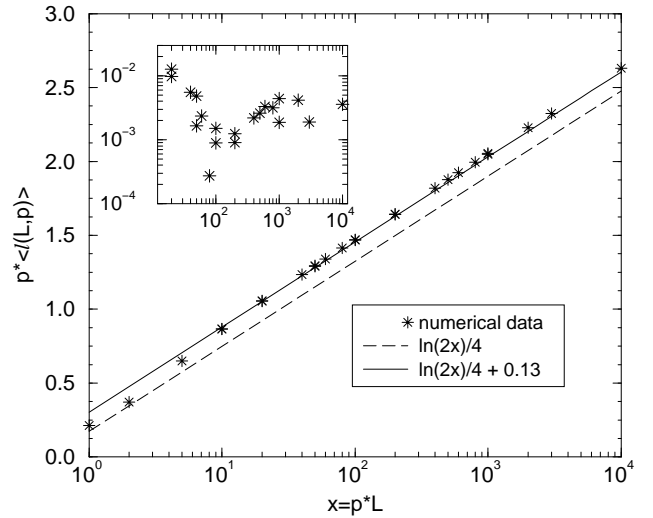}}
\caption{Numerically computed average minimal distance between two
         randomly chosen nodes, $\lb$, (stars) using p-values ranging
         from $10^{-4}$ to $10^{-3}$ and system sizes $L$ from $10^4$
         to $10^7$.  Dashed line: mean-field result $\lb_{\mathrm
         mft}/L = \log(2x) / (4x)$. Solid line: our fitting form
         $\lb/L = \lb_{\mathrm mft}/L + 0.13/x$. The inset shows the
         absolute relative error between the numerical data and our
         fitting form.}
\label{fig:lbar_new}
\end{figure}

To determine $\la m \ra$, we use the exact relation, which was proved
in \cite{kulkarni00} and is valid for a general degree distribution,
\begin{equation}
\lb = \la m \ra \left(1+\frac{k}{L}\right) + \la m^2\ra ~\frac{k}{L},
\end{equation}
together with the definition $\mu_2 = \la m^2 \ra - \la m \ra ^2$.
Thus, once we have fitting forms for $\lb$ and $\mu_2$, we can
determine $ \la m \ra $. From our simulations, we have already
determined $\mu_2$ (see above). As for $\lb$, we find numerically
that
\begin{equation}
\frac{k\lb}{L} = \frac{\log(2x)}{4x} + \frac{\alpha}{x}\label{eq:lfit},
\end{equation}
with $\alpha = 0.13$. This numerical estimate agrees well with the
mathematically rigorous result derived in Ref. \cite{barbour00}. In
Fig.\ \ref{fig:lbar_new}, we show a comparison between the fitting
form (\ref{eq:lfit}), the numerically obtained values for $\lb$, and
the mean field result, $k \lb_{\mathrm mft}/L \rightarrow \log(2x) /
(4x)$. As the inset shows, Eq.\ (\ref{eq:lfit}) gives $\lb$ with a
relative error of less than $1\%$. Keeping only terms through ${\cal
O} (1/x)$, and recalling that $\mu_2/L^2 \sim 1/x^2$, we find that
$\la m \ra \approx \lb$. Hence, $ \la m \ra $ is given by
\begin{equation}
\langle m \rangle = \frac{L}{k}\left(\frac{\log(2x)}{4x} +
     \frac{0.13}{x}\right). \label{eq:mav1}
\end{equation}

The three equations (\ref{eq:scale}), (\ref{eq:gfit}), and
(\ref{eq:mav1}) provide a full functional form for $f(m)$ in the
asymptotic scaling regime for a SWN with degree distribution ${\cal
D}_q$. Note the striking pattern of successive simplifications during
our analysis of the probability functions. Initially, $P(m|n)$
depended on the {\em four} variables $m$, $n$, $p$, and $N$. The exact
results obtained in Ref.\ \cite{kulkarni00} reduced this dependence to
a
\noindent
function of three variables $f(m;N,p)$. Furthermore, the scaling
properties discussed in \cite{kulkarni00} imply that, for $p \ll 1$,
$f(m;N,p) = h(m/N,pN)/N$, a function of only two variables. Finally,
we have shown in the present work that in the limit $x\gg 1$, the
small-world network is characterized by a function of a {\em single}
variable, $g(y)$. This functional form can be used to obtain detailed,
accurate, structural information about networks in the small-world
regime {\em without} having to perform computationally expensive
simulations.

The scaling forms for $f(m;N,p)$ can also be used to gain insight into
{\em dynamics} on SWN's. As an example, we consider the dynamics of a
random walker on SWN's, a problem which has been studied by several
authors \cite{monasson99,jespersen00,pandit01}. Here, we focus on the
mean-square displacement, $\la r^2 \ra$ of a random walker on a SWN,
as a function of time\cite{rw}. At long times, $\la r^2 \ra$ must
saturate to the finite value $\lbsq(L,p)$, since, in the long-time
limit, each node of the network has equal probability of being
occupied by the random walker. The other length scale determining the
behavior of the random walker is $\xi = 1/p$, the average distance the
walker travels to reach a shortcut. Note that in the scaling limit,
$\lbsq(L,p)/L^2$, and consequently $\xi^2/\lbsq(L,p)$, are functions
of $x$ only. Correspondingly we can write down the scaling ansatz:
\begin{equation}
\la r^2 \ra = \lbsq~ R\left(\frac{t}{\lbsq};x\right).
\label{eq:diffn}
\end{equation}
Because of the diffusive behavior for small $y$ and saturation for
large $y$, we expect $R(y) = y$ for $y \ll \xi^2/\lbsq$, and $R(y) =
1$ for $y \gg 1$. We have numerically confirmed the scaling collapse
implied by Eq.\ (\ref{eq:diffn}) for a wide range of 
\begin{figure}[tb]
\epsfysize=7cm
\centerline{\epsffile{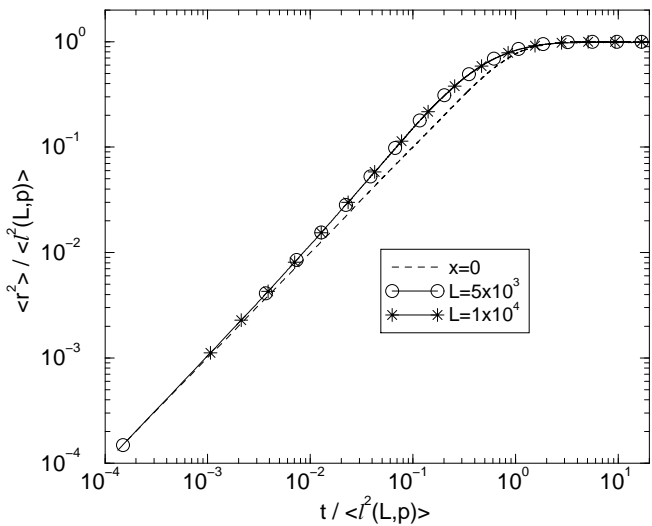}}
\caption{The mean-square displacement, $\la r^2 \ra$, for a random
         walker on SWN's with $x=100$ (upper curve), using $L=5\times
         10^3$ (circles and full line), and $1\times 10^4$ (stars and
         full line).  By contrast we also show $\la r^2 \ra$ for a
         regular network ($x=0$; dashed line).}
\label{fig:RW}
\end{figure}\noindent
$x$ values, and an example is shown in Fig.\ \ref{fig:RW}. Note
that we can readily calculate the length scale $\lbsq$ using our
functional form for $f(m)$, illustrating how knowledge of the
underlying structure helps in understanding the dynamical properties.

In summary, we have given exact results which are valid for SWN's with
an arbitrary degree distribution superimposed on a regular lattice.
We have also shown that in the small-world limit ($x \gg 1$), the
structure of the original SWN's can be characterized by a single
function, $g(y)$, and the mean value, $\la m \ra$, and we provide
empirical fitting forms for both quantities. In addition, we have
shown that the mean-square displacement of a random walker exhibits
scaling properties which follow from those of the underlying SWN. The
structural aspects of SWN's discussed in this work should be useful in
understanding many other properties of dynamics on SWN's.

The authors would like to thank S. V. Barabash for helpful
comments. This work has been supported by NASA through Grant NCC8-152
and NSF through Grant DMR01-04987 (EA and DS). RVK would like to
acknowledge the support of the US Department of Energy, Office of
Science, Division of Materials Research. Computational support was
provided by the Ohio Supercomputer Center, and the Norwegian
University of Science and Technology (NTNU). EA also thanks the Santa
Fe Institute (SFI) for their hospitality.


\begin{references}
\bibitem{fmnote} We do not indicate the dependence on the network
parameters $L$ and $p$, but include it later wherever needed for
clarity.
\bibitem{rw} Additional properties of the random walk on SWN's will be
discussed elsewhere.
\end{references}
\end{document}